\documentclass{icrc}

\usepackage{times}
 \usepackage{graphicx} 

\begin{document}

\title{Lateral Distribution Studies for the PACT Experiment}
\author[1]{P.R.Vishwanath}
\affil[1]{Tata Institute of Fundamental Research,Colaba\\
Mumbai,400005, India}

\correspondence{vishwa@tifr.res.in}

\runninghead{P.R.Vishwanath: Lateral Distribution Studies for PACT}
\firstpage{1}
\pubyear{2001}


\maketitle

\begin{abstract}
Monte Carlo calculations have been done to study the possibilities 
of using different spatial distributions, local fluctuations and the 
number of Cerenkov photons in gamma ray showers to increase the 
sensitivity of the PACT experiment. After matching the trigger rates in 
data with the predictions of simulations, the energy threshold of the 
array for gamma rays is found to be about 900 GeV. It is shown that  a 
lateral distribution parameter $\beta$ can increase the signal to noise ratio 
in the experiment. A typical season of 15 nights should be able to 
conclusively establish any source with Crab level fluxes. Two other 
possible parameters for increase of sensitivity are also discussed.  
Preliminary results on the application of the $\beta$ parameter to the data 
from PACT experiment are discussed.
\end{abstract}

\section{Introduction}
It is well known that lateral distribution of gamma ray showers is 
significantly different from those of proton showers.  
Cerenkov photon 
densities in gamma ray showers are almost flat from the core upto 110 
meters with a hump region between 110 meters and 130 meters whereas 
proton showers show a monotonically decreasing density with core distance.  
The Tata group has been  advocating the use of lateral distribution as a 
discriminating factor between protons and gamma ray events [\cite{Rao88}] 
[\cite{Vishwa93}]. It has been used for the pulsar data taken in 1992-1994 
and it was shown that the events in the main pulse region for Crab and the 
preferred GeV emission phase region for Geminga display gamma ray like 
characteristics [\cite{Vishwa97}]. Calculations have been done for the 
newly commissioned PACT [\cite{Bha00}] which show that the use of 
lateral distribution(LD) parameter can 
increase the S/N ratio. The recent data taken with the PACT array on Crab is 
used to check some of the predictions of these calculations.

\section{Simulations for the array}


Monte  Carlo simulations for the new PACT array were done  to 
understand the performance  of the experiment and have been discussed in 
detail in other papers of the conference.  Events were generated with  shower 
cores uniformly distributed inside  a circle of 300  meters radius from the 
centre of the array.  The energy for the event was picked from the energy 
spectrum of the cosmic rays  (gamma rays) with a slope of 1.66  (-1.4) for 
protons (gamma rays). The energy range for these calculations was 500 GeV 
 20 TeV. Chitnis and Bhat [\cite{Chi99}] have done extensive 
simulations in the recent years for atmospheric Cerenkov production from 
protons and gamma rays. The Lateral Distribution curves given by 
them for both protons and gammas were parametrized to cover the entire 
energy range. The number of Cerenkov photons at each mirror was obtained 
from this parametrization.  These were converted to photoelectrons using a 
mean quantum efficiency and night sky background photoelectrons were 
added to these. Thus, the number of photoelectrons at each mirror of the 
PACT array  was written out for about  0.2 million proton events and 0.1 
million gamma ray events. The trigger requirement in the experiment was 
any four out of six telescope coincidence  per sector. It was found that the 
experimental trigger rate for sector of about 4 Hz corresponds to about 55 
photoelectrons for an individual telescope which translates to 900 GeV 
energy threshold for gamma rays and 2.2 TeV for protons.




\section{Lateral distibutions}
While several methods were used by us in the past to exploit the 
flatness concept, a single parameter $\beta$ is used to exploit the full lateral 
distribution curve. $\beta$ is defined as ( max pulse height /mean  pulse height 
of the rest  -1). It is intuitively obvious that $\beta$  should have a small value for 
gamma ray showers when the core distances are not far from the center 
of the array. At larger distances, $\beta$ will have a larger value for gamma 
ray showers because few detectors are hit either by the flat region or the 
hump whereas the other detectors will respond to the falling part of the LD 
curve.  When median values of $\beta$ were calculated, it was  seen that 
median $\beta$ increases from 0.05  ( 0.2) at $<$ 50 meters to 0.45 (.15) at 200 
meters from the core for gamma and proton showers respectively. Further, it 
was also seen that higher energy events have a larger median $\beta$ for 
gamma ray showers. 

Monte Carlo generated events which fulfilled threshold criteria were 
binned according to their $\beta$ values and the number of gamma and proton 
showers triggering the array  as a function of $\beta$ were obtained. Fig 1(a) 
shows the fraction of events as s function of $\beta$.  As expected, the fraction 
of gamma ray events is higher at near  and far core distances compared to 
the fraction of proton events. Fig 1(b) shows the signal strength for a crab flux 
of 2.5/minute with an exposure of 55 hours. It is seen that both low and high 
$\beta$ regions can be exploited to increase the signal to noise ratio. 

\section{Comparison with Data}
The data taken on Crab with the PACT array during the months of  
2000 October and November  totaling 2301 minutes were used to see 
whether the above mentioned LD parameter could be used to extract the 
signal from the source. ADC counts from mirrors were converted to pulse 
heights and normalized such that all mirrors have the same mean pulse 
height for both source and background. From the individual mirror 
responses, the total pulse height from each telescope was obtained and $\beta$ 
parameter calculated and binned.  Fig. 1(c) shows the normalized ratio of 
source to background events as a function of $\beta$.  There is almost exact 
similarity between the two figures, the one obtained from Monte Carlo and 
the other from data. However, absolute values of the $\beta$ in data are higher 
than in Monte Carlo. One can see that there is an excess of events at very 
low  and very high $\beta$ as predicted by the simulations. The excess from the 
source amounts to 3.63 $\pm$ .38 per minute. The arrival direction for each 
event was calculated as described  in the papers of the PACT experiment in 
this conference. From the space angle differences, it was found that that on-
axis showers show a higher fraction of source showers at low $\beta$, thus 
confirming the gamma ray nature of the events at low $\beta$. However, the 
events at large $\beta$ did not show any preference for on-axis showers

\section{Other Parameters}
According to the simulations of Chitnis and Bhat[\cite{Chi99}], while density 
fluctuations for both types of showers are core distance dependent, the 
amount of fluctuation is much less in gamma ray showers: for example, at 
100 meters the ratio of standard deviation to mean photon density is 2\% 
and 20\% for gamma ray and proton showers respectively. While this has 
been the input to the Monte Carlo for fluctuating the Cerenkov photons, it is 
of interest to see its effect on the events which trigger the array.  By using 
the photoelectron number at each mirror, the fractional error on the mean 
was computed for each telescope. By using the information from all 
telescopes, the overall mean of such a quantity (FEML) for the event and 
further for all Monte Carlo events was calculated. It was found that the fact 
that gamma ray events have a much smaller local fluctutation did translate to 
an overall mean FEML of 0.525 and 0.619 for gammas and protons 
respectively. The PACT array with information from as many as 175 mirrors 
is ideally suited to use this difference to increase the sensitivity. It should 
however be noted that the energy threshold for use of this parameter is 
significantly high, almost 1800 GeV for gamma rays since most showers 
will not have information from all mirrors. 
Another parameter of interest is the amount of Cerenkov light per 
telescope per event. It can be seen from the lateral distribution curves that 
the number of Cerenkov photons is roughly same  for gamma showers of 
energy E and for photon showers of energy 2E. When a sample of events has 
both protons and gamma rays, the mean amount of light per event in that 
sample will depend on the amount of gamma ray admixture. For a Crab 
spectrum of slope -1.4 ( as was used for these Monte Carlo calculations), the 
mean Cerenkov light for gamma rays triggering the telescope was higher 
than for proton events by 30\%. If there are gamma rays from pulsars, then 
the pulsar data should provide a check of this difference between gamma 
and proton showers. This is exactly what was seen in the analysis of the data 
from the interim array  ( 8 banks , each of area 2.5$m^{2}$ ) runs at 
Pachmarhi during 1992-1994. The ADC information was used to get the 
Average Pulse Height per bank (APHB) per event as function of the pulsar 
phase. For the Crab pulsar data, the  difference in APHB between that  in the 
phase region 0-0.5 to the rest of the phase plot was 1.26 $\pm$ 0.26, a 5$\sigma$ 
effect. Similar, but slightly less significant , effect was seen in the 
comparison for the Geminga data. It should be noted that the right place to 
look for such effects is in the data from pulsars ( assuming there is pulsed 
emission at some phases) since all data would have been treated in the same 
manner ; it might be difficult to see such effects when the source region and 
background regions are entirely different.

\section{Conclusions}
It should be noted that the Q values (Quality factors) are not 
necessarily very high with the usage of the $\beta$ parameter. Fluctuations 
degrade the Q values which should have been ideally much higher. 
Nevertheless, it is still an important sensitivity raising parameter for steady 
emitters like the Crab nebula. Fig 1(b) which shows the expectations for a 55 
hour data sample is essentially for a typical 10-12  night run on a steady 
source  with Crab like flux levels. Further, some of the deficiencies of the 
Monte Carlo simulations have to be noted. The photon densities were picked 
from fluctuations imposed on the LD curve for the particular energy. The 
densities for higher energy showers are picked from the extrapolations of 
the  published LD curves at rather low energies. Thus, it is not apparent how 
well the real LD at the very high energies correspond to these extrapolated 
values. Further, the uncertainty in the attenuation factors, the reflectivity of the 
mirrors etc would affect the quantities obtained from the Monte Carlo 
calculations.





%
\begin{figure}[t]
\includegraphics[width=8.3cm]{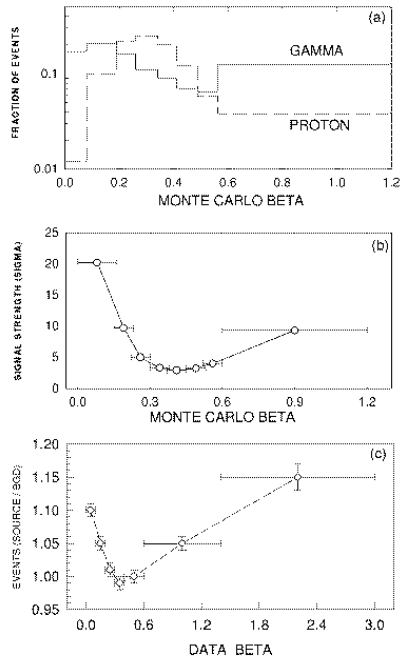} 
\caption{(a) Fraction of Triggered showers as a function of $\beta$ \\
(b) Signal Strength v/s $\beta$ for a Crab like source with exposure of
55 hours. (c) Ratio of Source to Background Events v/s $\beta$ for Crab data
of 2000.}
\end{figure}

%




\begin{acknowledgements}
It is a pleasure to thank Profs P.N.Bhat, \\
B.S.Acharya, 
Dr V.Chitnis 
and 
Mr.P.Majumdar for fruitful discussions.

\end{acknowledgements}





\begin{thebibliography}{99}

\bibitem[Bhat(2000)]{Bha00}
Bhat P.N. {\it et al}, Bull. Astro. Soc. India, 28, 45(2000)
\bibitem[Chitnis(1999)]{Chi99}
Chitnis V.R. and Bhat P.N., Astroparticle Physics, 12, 45(1999)
\bibitem[Rao(1988)]{Rao88}
Rao M.V.S. and Sinha S., Jour of Phys,G, 14, 811(1988)
\bibitem[Vishwanath(1993)]{Vishwa93}
Vishwanath P.R. {\it et al}, Towards a Major Atmospheric Cerenkov Detector,
II(Ed: R.Lamb, Iowa),(1993)
\bibitem[Vishwanath(1997)]{Vishwa97}
Vishwanath P.R., High Energy Astronomy and Astrophysics (Eds P.C.Agrawal
and P.R.Vishwanath, University Press), 204(1997) 
\end{thebibliography}
\end{document}